\documentclass[a4paper]{article}
\usepackage{flushend}  
\usepackage{blindtext, graphicx}
\usepackage{bm}
\usepackage{booktabs}
\usepackage{subfig}
\usepackage[font=footnotesize]{caption}
\usepackage{hyperref}
\usepackage{xspace}
\usepackage[binary-units]{siunitx}

\usepackage{caption}
\usepackage{localmath}

\usepackage{INTERSPEECH2022}

\usepackage{IEEEtrantools}

\graphicspath{{figures/}}

\title{End-to-End Multi-speaker ASR with Independent Vector Analysis}
\name{Robin Scheibler$^1$, Wangyou Zhang$^2$, Xuankai Chang$^3$, Shinji Watanabe$^3$, Yanmin Qian$^2$}
\address{
  $^1$LINE Corporation, Tokyo, Japan\\
  $^2$Shanghai Jiao Tong University, Shanghai, China\\
  $^3$Carnegie Mellon University, Pittsburgh, USA}
\email{robin.scheibler@linecorp.com}

\begin{document}
\bstctlcite{IEEEexample:BSTcontrol} 

\maketitle
\begin{abstract}
  We develop an end-to-end system for multi-channel, multi-speaker automatic speech recognition.
  We propose a frontend for joint source separation and dereverberation based on the independent vector analysis (IVA) paradigm.
  It uses the fast and stable iterative source steering algorithm together with a neural source model.
  The parameters from the ASR module and the neural source model are optimized jointly from the ASR loss itself.
  We demonstrate competitive performance with previous systems using neural beamforming frontends.
  First, we explore the trade-offs when using various number of channels for training and testing.
  Second, we demonstrate that the proposed IVA frontend performs well on noisy data, even when trained on clean mixtures only.
  Furthermore, it extends without retraining to the separation of more speakers, which is demonstrated on mixtures of three and four speakers.
\end{abstract}

\section{Introduction}

Automatic speech recognition (ASR) technology provides a natural interface for human-to-machine communication~\cite{yuAutomaticSpeechRecognition2015}.
Despite tremendous progress in the last decade, ASR systems are still severely challenged by reverberation, overlapped speech, and noise~\cite{haeb-umbachFarFieldAutomaticSpeech2021}.
Microphone arrays are a powerful tool to fight these degradations.
In particular, linear spatial filtering, i.e., beamforming, has been shown to reliably decrease the word error rate (WER) of ASR systems~\cite{haeb-umbachFarFieldAutomaticSpeech2021}.
While optimal beamforming formulations exist~\cite{VanTrees_optimum_2002}, e.g. the famous minimum variance distortionless response (MVDR) beamformer, their use has been traditionally limited by the difficulty of estimating the target and noise statistics.
These limitations have been recently practically solved by using trained neural networks to estimate these statistics~\cite{erdoganImprovedMVDRBeamforming2016}.
The resulting neural beamformers are highly effective~\cite{heymannNeuralNetworkBased2016}.
However, training these networks requires a large amount of parallel speech data, e.g. reverberant mixtures and the isolated anechoic sources they contain.
Such data is notoriously difficult to collect.
Instead, most works rely on simulation~\cite{habetsGeneratingNonstationary2008,Scheibler:2018di}.
However, the simulation is often insufficient, and some fully unsupervised approaches have been proposed~\cite{drudeUnsupervisedTrainingDeep2019,togamiUnsupervisedTrainingDeep2020}.

The situation for ASR systems is much different since transcripts of actual recordings may be collected by skilled annotators~\cite{yuAutomaticSpeechRecognition2015}.
Indeed, a large amount of annotated speech data has been collected for academic and commercial purposes~\cite{pratapMLSLargeScaleMultilingual2020}.
One can thus bypass the necessity of parallel speech data by concatenating enhancement and ASR systems, and training directly from the ASR loss~\cite{heymannBeamnetEndtoendTraining2017,Unified-Ochiai2017}.
Building on this approach, an end-to-end (E2E) paradigm for multi-channel, multi-speaker ASR called MIMO-speech~\cite{changMIMOSpeechEndtoendMultichannel2019} has been proposed.
This approach has demonstrated not only competitive ASR, but also decent separation performance, trained from the ASR loss only.
It has been extended to include several advanced joint dereverberation and beamforming methods~\cite{zhangEndtoEndFarFieldSpeech2020,zhangEndtoEndDereverberationBeamforming2021}.
Despite all these progresses, the challenge of domain mismatch remains.
Neural beamformers for separation typically rely on estimating multiple sources, usually two, from a single spectrogram.
If the test data is sufficiently different from the input data, this stage may fail, impeding the beamforming performance.

An alternative line of research builds upon independent vector analysis (IVA)~\cite{kimIndependentVectorAnalysis2006a,hiroeSolutionPermutationProblem2006}.
In addition to a statistical model of the sources, their mutual statistical independence is leveraged to help the separation.
Vanilla IVA is a blind method, requiring no training data, that can be solved iteratively~\cite{onoStableFastUpdate2011,scheiblerSurrogateSourceModel2021}, and rivals sophisticated neural beamformers~\cite{boeddeker_comparison_2021}.
Extensions to joint dereverberation and separation have been proposed~\cite{ilrma-t}.
In particular, time-decorrelation iterative source steering (T-ISS)~\cite{nakashimaJointDereverberationSeparation2021} is a stable and fast algorithm that avoids matrix inversion.
Recently, combining IVA with a neural source model has attracted attention~\cite{idlma,kameoka_supervised_2019}.
One particular approach proposes to train a neural source model end-to-end through T-ISS~\cite{scheiblerSurrogateSourceModel2021,saijo_tiss_2021}.
Unlike in neural beamforming, the network models a single source.
This allows to maintain the high performance of neural beamforming while being agnostic to the number of sources and channels, and robust to a fair amount of data mismatch~\cite{saijo_tiss_2021}.


In this work, we investigate the use of a T-ISS frontend for MIMO-speech E2E ASR.
We extend T-ISS to the overdetermined case, where more channels than sources are present~\cite{duComputationallyEfficientOverdeterminedBlind2021}
We implement our model in ESPnet~\cite{Watanabe:2018gy} with an E2E transformer-based ASR backend~\cite{karitaComparativeStudyTransformer2019}.
The whole system is trained E2E by joint CTC/attention loss~\cite{Joint-Kim2017}.
We use variations of the spatialized WSJ1~\cite{changMIMOSpeechEndtoendMultichannel2019} dataset, clean, and corrupted with several kinds of noise.
We explore how the number of channels at training affects test performance (spoiler: more is better).
We demonstrate the robustness of T-ISS to mismatch with training in terms of noise and number of sources.

\section{Background}

We use the following notation. Bold lower and upper case letters are for vector and matrices, respectively.
Furthermore, $\mA^\top$ and $\mA^\H$ denote the transpose and conjugate transpose, respectively, of matrix $\mA$.
The norm of vector $\vv$ is $\|\vv\| = (\vv^\H \vv)^{\half}$.

\begin{figure}
  \centering
  \includegraphics[width=\linewidth]{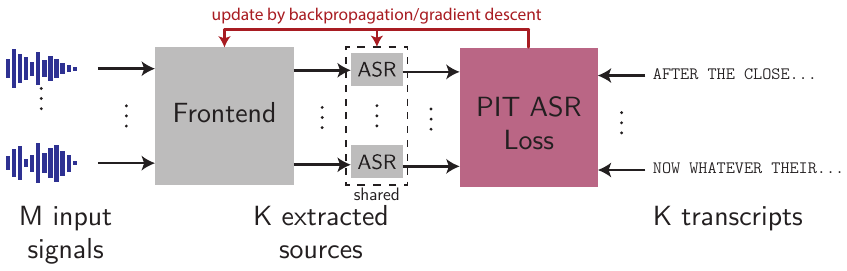}
  \caption{Illustration of the MIMO-speech E2E system for $K$ source and $M$ microphones.}
  \flabel{e2e_asr_system}
\end{figure}

\subsection{MIMO-Speech and Its Extensions}

We are concerned with multi-channel, multi-speaker ASR systems.
The MIMO-Speech method~\cite{changMIMOSpeechEndtoendMultichannel2019,changEndToEndMultiSpeaker2020} proposes a fully end-to-end framework that jointly optimizes the entire system with only the final ASR criterion.
The model consists of a beamforming-based frontend for speech separation and an E2E ASR backend.
It takes $M$ channels, as inputs, and outputs $K$ text hypotheses, corresponding to $K$ concurrent speakers.
First, the frontend extracts $K$ source signals from the input mixture.
Second, each of the extracted signal is processed by the same E2E ASR backend in parallel.
This produces $K$ text hypotheses that are evaluated against the $K$ reference transcripts with a permutation invariant training (PIT) loss~\cite{kolbaekMultitalkerSpeechSeparation2017}.
The process is illustrated in \ffref{e2e_asr_system}.

\subsection{Multichannel Speech Separation and Dereverberation}

Physically, the signals from the $K$ sources propagate and reflect on the walls of a room, and mix additively with various amplitudes and time delays at the $M$ microphones.
This process can be approximated in the short-time Fourier transform (STFT) domain as follows,
\begin{align}
  \vx_{fn} = \mA_f \vs_{fn} + \mZ_f \bar{\vx}_{fn}  + \vb_{fn}, \quad \in \C^M,
  \elabel{signal_model}
\end{align}
where $\vs_{fn} \in \C^K$ is a vector containing the source signals.
The matrix $\mA_f\in \C^{M\times K}$ contains the transfer functions from sources to microphones in its entries.
The term $\bar{\vx}_{fn} = \begin{bmatrix} \vx_{f,n-D}^\top,\ldots,\vx_{f,n-D-L}^\top\end{bmatrix}^\top$ accounts for reflections from the room mixed by the matrix $\mZ_f \in \C^{M \times ML}$.
The reverberation length is given by $L$, and $D$ is a delay necessary due to the overlap of the STFT.
Finally, $\vb_{fn}\in \C^M$ is a noise term
The indices $f$ and $n$ are for frequency bins and STFT frames, and run from 1 to $F$ and $N$, respectively.
The role of the frontend is to reduce the second and third terms, as much as possible, and invert $\mA_f$, if feasible.

\subsubsection{Neural Dereverberation and Beamforming}

Conventionally, dereverberation and beamforming are done in distinct steps.
While several dereverberation methods exist, WPE~\cite{nakataniSpeechDereverberationBased2010} has been widely adopted for ASR.
Ignoring the noise term $\vb_{fn}$, the dereverberated mixture $\mA_f \vs_{fn}$ can be obtained if we know $\mZ_f$.
Provided with a neural network producing a mask $u^{\text{WPE}}_{fn}$ hiding the target signal from the spectrogram, the dereverberation filters are given by the minimizer of
$\sum_n u_{fn}^{\text{WPE}} \| \vx_{fn} - \mZ_f \bar{\vx}_{fn}\|^2$.
The beamforming filters are computed from the spatial covariance matrices of target speech and noise.
Since these are typically not available, a neural network is trained to estimate masks that extract these signals power from a spectrogram.
Let $u_{fn}$ be one of these masks.
Then, the corresponding spatial covariance matrix is $\mPhi_f = \nicefrac{1}{N}\sum_n u_{fn} \tilde{\vx}_{fn} \tilde{\vx}_{fn}^\H$, where $\tilde{\vx}_{fn}$ is the dereverberated input signal.
For the separation of multiple sources, the neural network is trained to output masks for each of the sources.
Several ways of combining or sharing masks between steps have been proposed and give rise to different neural beamformers such as MVDR, WPD~\cite{nakataniUnifiedConvolutionalBeamformer2019}, and wMPDR~\cite{nakataniJointlyOptimalDenoising2020}.
See~\cite{zhangEndtoEndFarFieldSpeech2020} for the details.
We emphasize here that after the masks are estimated, the beamforming filters are estimated independently.
Thus, if the mask estimation fails, it cannot be corrected during the filter computation step.

\subsubsection{Independence-based Dereverberation and Separation}
\seclabel{background:t-iss}

This approach builds upon independent vector analysis (IVA)~\cite{kimIndependentVectorAnalysis2006a,hiroeSolutionPermutationProblem2006}.
Unlike, the neural beamforming approach, the foundational hypothesis of IVA is that multiple statistically independent sources are present.
The approach has been extended to jointly optimize for dereverberation~\cite{ilrma-t,nakashimaJointDereverberationSeparation2021} and include a trainable neural source model~\cite{scheiblerSurrogateSourceModel2021,saijo_tiss_2021}.
We define the demixing and dereverberation matrices as $\mW_f \in \C^{M\times M}$ and $\mU_f\in \C^{M \times ML}$, respectively.
For convenience, we concatenate them into a unified dereverberation and separation matrix $\mP_f = [\mW_f\ \mU_f]$.
Further let $\tilde{\vx}_{fn}$ be the concatenation of $\vx_{fn}$ and $\bar{\vx}_{fn}$.
The $k$th row of $\mP_f$, i.e. $\vp_{kf}^\H$, extracts the $k$th target as $y_{kfn} = \vp_{kf}^\H \tilde{\vx}_{fn}$.
Maximum likelihood estimation of $\mP_f$ can be done by iteratively minimizing the cost function
\begin{multline}
  \calL(\mP_f) =  \sum_{kfn} u_{fn}(\hat{\mY}_k) | \vp_{kf}^\H \tilde{\vx}_{fn} |^2 - 2\log |\det \mW_f| 
  \elabel{t-iss-cost}
\end{multline}
where $\vp_{kf}^\H$ is the $k$th row of $\mP_f$.
The $F\times N$ matrix $\hat{\mY}_k$ is the current estimate of source $k$ with entries $(\hat{\mY}_k)_{fn} = \hat{y}_{kfn}$.
The cost function \eref{t-iss-cost} is derived from the likelihood function of the observed data $\vx_{fn}$~\cite{ilrma-t}.
The change of variable from $\vx_{fn}$ to $\mY_k$ allows to work on the source signals, rather than the mixture, but introduces the log-determinant term.
Then, since the sources are independent, the joint probability density function (pdf) is the product of the marginals.
Finally, we assume the log-pdf may be majorized by that of the Normal distribution to obtain \eref{t-iss-cost}~\cite{onoStableFastUpdate2011}.
The function $u_{fn}(\mY)$ is derived from the majorization step, and guarantees decrease of~\eref{t-iss-cost}.
In~\cite{scheiblerSurrogateSourceModel2021,saijo_tiss_2021}, this exact derivation is abandoned, together with the guarantees, and $u_{fn}(\mY)$ is replaced by a trainable neural network.

While minimization of \eref{t-iss-cost} does not have a closed-form solution, algorithms to efficiently decrease its value exist~\cite{ilrma-t,nakashimaJointDereverberationSeparation2021}.
T-ISS~\cite{nakashimaJointDereverberationSeparation2021} is particularly suitable for use in E2E training because of low computational cost and lack of matrix inversion.
The algorithm proceeds by finding a sequence of $\ell = 1,\ldots, M(L+1)$ optimal rank-1 updates of the form
\begin{align}
  \mP_f \gets \mP_f - \vv_\ell \vp_{\ell f}^\H, \quad \vv_\ell = \underset{\vv \in \C^M}{\arg\min}\ \calL(\mP_f - \vv \vp_{\ell f}^\H).
  \elabel{t-iss-update}
\end{align}
For $\ell > M$, we define $\vp_{\ell f} = \ve_\ell$, i.e., the vector with all zeros but a one at position $\ell$.
The closed form solution for $\vv_\ell$ in \eref{t-iss-update} is
\begin{align}
    (\vv_{\ell})_q = 
    \begin{cases}
      1 - \left(\sum_{n} \frac{u_{fn}(\bm{Y}_\ell)}{N}|y_{\ell fn}|^2  \right)^{-\half}, & \text{if $q = \ell$,} \smallskip \\
      \frac{\sum_{n} u_{fn}(\bm{Y}_q) y_{qfn}y^{*}_{\ell f n}}{\sum_{n} u_{f n}(\bm{Y}_q) |y_{\ell f n}|^2 }, & \text{else}.
    \end{cases}
  \elabel{t-iss-update-eq}
\end{align}
where $y_{qfn} = \vp_{qf}^\H \tilde{\vx}_{fn}$ for $q \leq M$, and $y_{qfn} = \ve_\ell^\top \tilde{\vx}_{fn}$ else.
In contrast to the neural beamformer in~\cite{zhangEndtoEndFarFieldSpeech2020,zhangEndtoEndDereverberationBeamforming2021}, spatial cues are taken into account when estimating the source masks.
Because the neural network models a single source, the algorithm is easily extended to different numbers of sources.
It was also shown to be robust to domain mismatch~\cite{saijo_tiss_2021}.

\section{Proposed End-to-end Architecture}

\begin{figure}
  \centering
  \includegraphics[width=\linewidth]{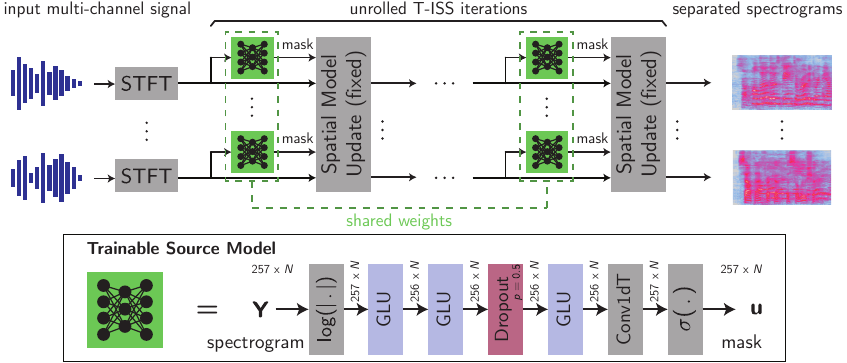}
  \caption{Block diagram of the proposed frontend and neural source model.}
  \flabel{overview_e2e_tiss_asr}
\end{figure}

Our proposed system builds upon the latest methodology of MIMO-Speech~\cite{zhangEndtoEndDereverberationBeamforming2021}.
We replace the WPD beamforming frontend by an IVA-based one that performs joint dereverberation and separation.
During training, multiple iterations of T-ISS are run to obtain the separation matrix.
The same single source mask network is used for all iterations and separation output.
The proposed frontend is illustrated in \ffref{overview_e2e_tiss_asr}.
The system is trained E2E from the ASR loss as illustrated in~\ffref{e2e_asr_system}.

\subsection{ASR Transformer Model}


We adopt the joint connectionist temporal classification (CTC)/attention-based encoder-decoder~\cite{Joint-Kim2017} as the ASR backend, which consists of four submodules: feature extraction, encoder, CTC, and attention-decoder.
For each separated stream $\hat{\mathbf{Y}}_k$ from the frontend, 80-dimensional log-Mel filterbank features $\mathbf{O}_k$ are firstly extracted via the feature extraction module $\operatorname{MVN-LMF}$.
The extracted feature is then fed into the ASR encoder to obtain hidden representations $\mathbf{H}_k$, which are used in both CTC and attention-decoder submodules for recognition.
The ASR procedure is summarized as follows:
\begin{align}
  \begin{array}{ll}
    \mathbf{O}_k = \operatorname{MVN-LMF}(\hat{\mathbf{Y}}_k), & 
    \mathbf{H}_k = \operatorname{Enc}(\mathbf{O}_k), \\
    \hat{\mathbf{R}}^{\text{(ctc)}}_k = \operatorname{CTC}(\mathbf{H}_k), &
    \hat{\mathbf{R}}^{\text{(dec)}}_k = \operatorname{AttentionDec}(\mathbf{H}_k),
  \end{array}
  \nonumber
\end{align}
where $\operatorname{MVN-LMF}$ denotes extracting log-Mel filterbank features and applying mean-variance normalization. $\hat{\mathbf{R}}^{\text{(ctc)}}_k$ and $\hat{\mathbf{R}}^{\text{(dec)}}_k$ are recognition results from CTC and attention-decoder submodules, respectively.
The ASR loss function is constructed based on multi-task learning:
\begin{align}
    \mathcal{L}_{asr} &= \alpha \mathcal{L}_{ctc} + (1 - \alpha) \mathcal{L}_{dec}, \elabel{loss}
\end{align}
where $0 < \alpha < 1$ is an interpolation factor.
Note that the permutation invariant training (PIT)~\cite{kolbaekMultitalkerSpeechSeparation2017} method is applied in the CTC submodule as in~\cite{changMIMOSpeechEndtoendMultichannel2019,changEndToEndMultiSpeaker2020} to solve the label permutation problem when processing multiple separated streams.

\subsection{Overdetermined T-ISS Frontend}

Our independence-based frontend is a new extension of the T-ISS~\cite{nakashimaJointDereverberationSeparation2021} algorithm, described in this sub-section, that can be used when more channels are available than there are sources.
When there are more channels than sources, i.e., $K < M$, the separation matrix $\mW_f$ is not square anymore and the algorithm of \sref{background:t-iss} is not sufficient anymore.
An extension of ISS to the overdetermined case has been proposed, but for separation only~\cite{duComputationallyEfficientOverdeterminedBlind2021}.
We extend it here to include dereverberation.

First, we write the overdetermined dereverberation and separation operation as a determined system, i.e., square,
\begin{align}
  \begin{bmatrix}
    \vy_{fn} \\ \vz_{fn} \\ \bar{\vx}_{fn}
  \end{bmatrix}
  = \begin{bmatrix}
    \mW_f & \mU_f \\
    \begin{bmatrix} \mJ_f & -\mI_{M-N} \end{bmatrix} & \vzero \\
    \vzero & \mI_{M L}
  \end{bmatrix}
  \begin{bmatrix} \vx_{fn} \\ \bar{\vx}_{fn} \end{bmatrix}.
  \elabel{prop:overdet_sys}
\end{align}
The separated target sources are $\vy_{fn}\in\C^K$, and $\vz_{fn}\in\C^{M-K}$ is a vector of background noise sources to make the system determined.
The top part contains $\mP_f = [\mW_f\ \mU_f]$, of \sref{background:t-iss}, but with $K$ rows, since we only wish to extract so many sources.
To complete the separation matrix, we add the strict minimum of parameters $\mJ_f\in \C^{M-K\times K}$.
The zeros on the right of the middle block reflect that we do not need to dereverberate the background noise.
Previous work~\cite{scheibler_mm_2020,ikeshitaOverdeterminedIndependentVector2020,togamiOverdeterminedSpeechSource2020a} has shown that a necessary condition for optimality is that the target sources and the noise vector be orthogonal, i.e., $\Expect{\vy_{fn} \vz_{fn}^\H} = \vzero$.
From this and \eref{prop:overdet_sys}, we obtain an equation for $\mJ$,
\begin{align}
  ((\mW_f \mR_f + \mU \overline{\mC}_f) \mE_1) \mJ^\H = ((\mW_f \mR_f + \mU \overline{\mC}_f) \mE_2),
  \elabel{background_update}
\end{align}
where $\mR_f = \Expect{\vx_{fn} \vx_{fn}^\H}$, $\overline{\mC}_f = \Expect{\bar{\vx}_{fn} \vx_{fn}^\H}$, and $\mE_1$, $\mE_2$ are of the appropriate shape, and such that $[\mE_1\ \mE_2] = \mI$.
Following the methodology of~\cite{duComputationallyEfficientOverdeterminedBlind2021}, we update $\mP_f = [\mW_f\ \mU_f]$ and $\mJ_f$ in two steps.
\begin{enumerate}
  \item We update $\mP_f = [\mW_f \mU_f]$ with \eref{t-iss-update} and \eref{t-iss-update-eq}, but limit the size of $\vv_\ell$ to $K$ to match the size of $\mW_f$, $\mU_f$ in~\eref{prop:overdet_sys}.
  \item Update $\mJ$ by solving~\eref{background_update}.
\end{enumerate}
The resulting algorithm maintains the low-complexity of T-ISS while allowing to use more channels for increased separation power.
We note that one matrix inverse is introduced in step~2.
However, the size of the matrix to invert is only $K\times K$, with e.g., $K=2$ for two sources.
Despite this small size, we observed some stability issues.
The system matrix is not Hermitian symmetric, and its eigenvalues not always positive.
Thus, straight diagonal loading, as in~\cite{zhangEndtoEndDereverberationBeamforming2021}, does not guarantee stability.
Our solution is to replace the $d\times d$ system $\mA \vx = \vb$ by
\begin{align}
  (\mA^\H\mD^{-1}\mA + \epsilon \mI) \vx =  \mA^\H \mD^{-1} \vb,
\end{align}
where $\mD$ is a diagonal matrix containing the square norms of the rows of $\mA$.
The system matrix is now guaranteed positive definite.
Clearly, if $\epsilon = 0$, the solution is the same as the original system.
Furthermore, normalizing the rows of $\mA$ with $\mD$ ensures the sum of the eigenvalues of the system matrix is $d$.
This allows a numerically sensible choice of $\epsilon$.



\section{Experiments}

We conducted several experiments to assess the performance of the proposed method.
We investigate the impact of the number of channels, iterations, and the presence of noise.

\subsection{Experimental Conditions}

We evaluate the proposed method on three datasets derived from the WSJ1 corpus~\cite{wsj1}.
For training and test on clean and noisy mixtures, we use \text{wsj1\_spatialized}~\cite{changMIMOSpeechEndtoendMultichannel2019}, and a similar dataset~\cite{scheiblerSurrogateSourceModel2021} that includes noise from the CHiME3 challenge~\cite{barkerThirdCHiMESpeech2015}.
These two datasets have 8 and 6 channels, respectively.
For noisy training, we concatenate the clean and noisy training sets.
The third dataset is used for mismatched testing and remixes the clean test mixtures with simulated diffuse noise~\cite{habetsGeneratingNonstationary2008} created from the TUT environmental sound database~\cite{Mesaros2018_DCASE} with SNR of \SIrange{5}{15}{\decibel}.
Unlike previous work~\cite{zhangEndtoEndFarFieldSpeech2020,zhangEndtoEndDereverberationBeamforming2021}, we did not do multi-condition training and used only the reverberant mixtures.
All the input speech is sampled at \SI{16}{\kilo\hertz}.
The STFT uses a \SI{25}{\milli\second} long Hann window with \SI{10}{\milli\second} shifts.
The FFT is zero padded to length \num{512} producing \num{257}-dimensional spectral feature vectors.
After the frontend, spectrograms are converted to 80-dimensional log Mel-filterbank features.
The training was conducted on an NVidia V100 graphical processing unit (GPU) with \SI{32}{\giga\byte} RAM.

All the models are implemented in ESPnet~\cite{Watanabe:2018gy} using the PyTorch~\cite{paszkePyTorchImperativeStyle2019} backend.
For the baseline, we use the WPD model described in~\cite{zhangEndtoEndFarFieldSpeech2020}.
It uses a bidirectional long-short term memory~(BLSTM) network with 600 cells in each direction followed by an output layer producing three masks per target speaker, i.e. 6 in our case.
WPE is configured with $L=5$ taps and delay $D=3$, and runs for single iteration.
%
The neural source model for T-ISS is the same as in~\cite{scheiblerSurrogateSourceModel2021}.
It has three convolutional layers, with batch-norm, max pooling, and GLU activations.
It has a 256 hidden dimension and dropout set to \num{0.2}.
For the models trained on clean data, we preprocessed the input with 5 iterations of AuxIVA-ISS with a non-trainable source model~\cite{scheiblerFastStableBlind2020}.
This is followed by 10 iterations of T-ISS with the neural source model.
After separation, the scale and phase are aligned to a reference channel by projection back~\cite{murataApproachBlindSource2001}.
As it did not seem very effective, we did not apply this preprocessing when training on the noisy dataset.
Instead we ran 15 iterations of T-ISS straight.
We used demixing matrix checkpointing~\cite{saijo_tiss_2021} to allow the model to fit on GPU during training.

We used the Adam optimizer with warmup set to 25000 and 50000 steps on the clean and noisy datasets, respectively, and initial learning rate of \num{1}.
The WPD baseline was always trained with maximum $M=8$ channels.
We trained multiple T-ISS models on the clean dataset using $M \in \{2, 4, 8\}$ channels.
On the noisy dataset, we only trained on $M=4$ channels due to time constraints.
At test time, the number of T-ISS iterations was adjusted to achieve better performance.
An external word-level recurrent neural network language model (RNNLM)~\cite{horiEndToEnd2018} is applied as shallow fusion in the decoding stage.

\begin{table}
  \centering
  \caption{Performance in terms of WER\,(\%) on the clean test set. Models are trained on clean data with $M$ channels. The four columns on the right are for different number of channels at test time.}
  \footnotesize
  \begin{tabular}{@{}lrrrrr@{}}
    \toprule
    Algorithm    &   $M$ & 2ch    & 4ch    & 6ch    & 8ch  \\
    \midrule
    Best in~\cite{zhangEndtoEndDereverberationBeamforming2021}$^\dagger$ & 2ch & 15.01 & --- & 9.02& --- \\
    \cmidrule{1-6}
    WPD &     8ch & 25.71 & 12.56 & 10.00 & 9.57  \\
    T-ISS &     2ch & \bf 13.71 & 23.40 & 28.88 & 31.46 \\
    T-ISS &     4ch & 20.57 & 10.37 & 10.86 & 11.16 \\
    T-ISS &     8ch & 25.71 & \bf 9.98  & \bf 9.08  & \bf 9.16  \\
      \bottomrule
       \multicolumn{6}{@{}l@{}}{\footnotesize $^\dagger$ Included for reference, training and parameters differ.} \\
  \end{tabular}
  \tlabel{eval_noiseless}
\end{table}

\subsection{Experimental Results}

\begin{table}
  \caption{Performance on matched and mismatched data.}
  \resizebox{\columnwidth}{!}{
  \begin{tabular}{@{}lr@{~/~}lr@{~/~}lrr|rrr@{}}
  \toprule
    Algo.   & \multicolumn{2}{c}{Train} &   \multicolumn{2}{c}{Test} 
                                                    &   WER     &  SIR &   SDR &   PESQ &   STOI \\
    \midrule                                                            
     WPD    &    8ch & clean    &    8ch & clean    &     9.57  &     13.9 & \bf  6.9 & \bf  1.88 & \bf 0.855 \\
     T-ISS  &    8ch & clean    &    8ch & clean    & \bf 9.16  & \bf 16.8 &      3.7 &      1.78 &     0.830 \\
 \midrule                                                               
     WPD    &    8ch & clean    &    6ch & noisy    &     17.12 &     12.3 & \bf  8.7 &      1.70 &     0.890 \\
     T-ISS  &    8ch & clean    &    6ch & noisy    & \bf 12.48 & \bf 15.6 &      6.2 & \bf  1.86 & \bf 0.913 \\
 \midrule                                                               
     WPD    &    8ch & noisy    &    6ch & noisy    & \bf 11.40 & \bf 14.7 & \bf 10.8 & \bf  1.79 &     0.918 \\
     T-ISS  &    4ch & noisy    &    6ch & noisy    &     12.18 &     14.4 &      7.3 &      1.77 & \bf 0.922 \\
 \midrule                                                               
     WPD    &    8ch & noisy    &    8ch & TUT      &     15.17 &     10.0 & \bf  5.2 & \bf  1.57 & \bf 0.816 \\
     T-ISS  &    4ch & noisy    &    8ch & TUT      &     23.56 &     11.4 &      1.5 &      1.42 &     0.741 \\
     T-ISS  &    8ch & clean    &    8ch & TUT      & \bf 14.55 & \bf 13.7 &      2.1 &      1.45 &     0.787 \\
 \bottomrule
  \end{tabular}%
  }
  \tlabel{match_mismatch}
\end{table}

\textbf{Effect of Number of Channels} \tref{eval_noiseless} reports the ASR evaluation results on the clean test set in terms of WER.
Each row represents a different trained model.
The performance with different numbers of channels at test time is reported in the four right-most columns.
We observed that using more channels at training pays off.
Models trained this way had lower WERs, even when testing with fewer channels.
There is however an exception for T-ISS where the behavior differed if trained with two channels, or more.
When trained on two-channel data, the performance was outstanding on two channels test data, even better than the best result from~\cite{zhangEndtoEndDereverberationBeamforming2021}, but did poorly with more channels.
While not reported due to space constraints, separation metrics increased with the number of channels.
This suggests that the ASR backend overfits the artefacts of the separation stage for two channels.
Similarly, T-ISS models trained on more channels performed poorly on the two-channel test set.
The best performing model was T-ISS trained on 8 channels.

\textbf{Mismatched Conditions} \tref{match_mismatch} reports the ASR performance under different training and test conditions.
When trained and tested on clean data, both frontends achieved under \SI{10}{\percent} WER, with T-ISS slightly better at \SI{9.16}{\percent}.
However, when trained on clean, but tested on noisy data, T-ISS significantly outperformed WPD by \SI{4.6}{\percent}.
When trained on noisy data, the performance of WPD recovered.
T-ISS did about \SI{0.7}{\percent} worse than WPD, but still a little better than in the mismatched condition.
Note that in this case, the noise was from the CHiME3 dataset both for training and testing.
We thus further tested on the mismatched noisy mixture dataset (labeled TUT in the table).
For WPD, the noisy training was effective at improving the robustness, and the WER did not increase as much as before.
The T-ISS model trained on noisy data performed much worse on the mismatched noise.
We conjecture that this may be due to the lack of a proper noise model, and overfitting to the noise in the source model.
It was the T-ISS model trained on noiseless data only that performed best here.
\tref{match_mismatch} also shows the regular separation metrics SDR, SIR, PESQ, and STOI.
T-ISS had consistently high SIR, but otherwise somewhat lower metrics.
The SDR in particular is much lower than that of WPD.
This suggests that it achieves good separation, but at the expense of more target degradation.


\textbf{Separation of 3 and 4 speakers} Even though the model was trained on two speaker mixtures, the T-ISS algorithm can be used to separate more, provided that sufficiently many channels are available.
We tested this on 3 and 4 speakers mixtures from the noisy dataset using 6 channels.
\tref{more_speakers} shows the results.
We note that the problem becomes much harder than in the two speakers case since the per-speaker SNR drops significantly.
Still, reasonable ASR performance was maintained in this challenging situation.


\begin{table}
  \centering
  \caption{Performance of T-ISS trained with two speakers on mixtures $K=3,4$ speakers.}
  \footnotesize
  \begin{tabular}{@{}rl@{~/~}llrrrr@{}}
    \toprule
    $K$ & \multicolumn{2}{c}{Train}   & WER    &   SDR &   SIR &   PESQ &   STOI \\
    \midrule
          3 & 8ch & clean        &  \bf   17.80 &     3.9 & \bf 10.2 & \bf 1.52 &      0.862   \\
            & 4ch & noisy        &        18.28 & \bf 4.8 &      9.9 &     1.51 &  \bf 0.870 \\
                    \cmidrule{1-8}
          4 & 8ch & clean        &        33.06 &     1.1 &      5.8 & \bf 1.34 &     0.792 \\
            & 4ch & noisy        &  \bf   30.66 & \bf 2.2 &  \bf 6.1 & \bf 1.34 & \bf 0.804 \\
    \bottomrule
  \end{tabular}
  
  \tlabel{more_speakers}
\end{table}

\section{Conclusions}

We have proposed the joint training of a MIMO-speech ASR system with an independent vector analysis frontend using the T-ISS algorithm.
T-ISS is an iterative procedure performing joint separation and dereverberation with the help of a neural source model.
We demonstrate that E2E training of this system, through the iterations, yields an ASR robust to data mismatch.
The T-ISS frontend trained on clean data only, did best, or at least well enough, on all our test sets.
In contrast, the neural beamformer baseline required noisy data in the training set in order to avoid a large performance drop.

Future work should concentrate on the inclusion of a noise model in T-ISS, e.g.~\cite{koldovskyOrthogonallyConstrainedExtractionIndependent2018}.
Multi-condition training and curriculum learning are also promising research avenues.

\clearpage


\bibliographystyle{IEEEtran}
\bibliography{refs}

\begin{thebibliography}{10}
\providecommand{\url}[1]{#1}
\csname url@samestyle\endcsname
\providecommand{\newblock}{\relax}
\providecommand{\bibinfo}[2]{#2}
\providecommand{\BIBentrySTDinterwordspacing}{\spaceskip=0pt\relax}
\providecommand{\BIBentryALTinterwordstretchfactor}{4}
\providecommand{\BIBentryALTinterwordspacing}{\spaceskip=\fontdimen2\font plus
\BIBentryALTinterwordstretchfactor\fontdimen3\font minus
  \fontdimen4\font\relax}
\providecommand{\BIBforeignlanguage}[2]{{%
\expandafter\ifx\csname l@#1\endcsname\relax
\typeout{** WARNING: IEEEtran.bst: No hyphenation pattern has been}%
\typeout{** loaded for the language `#1'. Using the pattern for}%
\typeout{** the default language instead.}%
\else
\language=\csname l@#1\endcsname
\fi
#2}}
\providecommand{\BIBdecl}{\relax}
\BIBdecl

\bibitem{yuAutomaticSpeechRecognition2015}
D.~Yu and L.~Deng, \emph{Automatic {Speech} {Recognition}}, 1st~ed., ser.
  Signals and {Communication} {Technology}.\hskip 1em plus 0.5em minus
  0.4em\relax London: Springer-Verlag, 2015.

\bibitem{haeb-umbachFarFieldAutomaticSpeech2021}
R.~Haeb-Umbach \emph{et~al.}, ``Far-field automatic speech recognition,''
  \emph{Proc. IEEE}, vol. 109, no.~2, pp. 124--148, Feb. 2021.

\bibitem{VanTrees_optimum_2002}
H.~L. Van~Trees, \emph{Optimum Waveform Estimation}.\hskip 1em plus 0.5em minus
  0.4em\relax John Wiley \& Sons, Ltd, 2002, ch.~6, pp. 428--709.

\bibitem{erdoganImprovedMVDRBeamforming2016}
H.~Erdogan \emph{et~al.}, ``Improved {MVDR} beamforming using single-channel
  mask prediction networks,'' in \emph{INTERSPEECH}, Sep. 2016, pp. 1981--1985.

\bibitem{heymannNeuralNetworkBased2016}
J.~Heymann \emph{et~al.}, ``Neural network based spectral mask estimation for
  acoustic beamforming,'' in \emph{ICASSP}, Mar. 2016, pp. 196--200.

\bibitem{habetsGeneratingNonstationary2008}
E.~A. Habets \emph{et~al.}, ``Generating nonstationary multisensor signals
  under a spatial coherence constraint,'' \emph{The Journal of the Acoustical
  Society of America}, vol. 124, no.~5, pp. 2911--2917, 2008.

\bibitem{Scheibler:2018di}
R.~Scheibler \emph{et~al.}, ``Pyroomacoustics: A {Python} package for audio
  room simulation and array processing algorithms,'' in \emph{ICASSP}, Apr.
  2018, pp. 351--355.

\bibitem{drudeUnsupervisedTrainingDeep2019}
L.~Drude \emph{et~al.}, ``Unsupervised training of a deep clustering model for
  multichannel blind source separation,'' in \emph{ICASSP}, May 2019, pp.
  695--699.

\bibitem{togamiUnsupervisedTrainingDeep2020}
M.~Togami \emph{et~al.}, ``Unsupervised training for deep speech source
  separation with {K}ullback-{L}eibler divergence based probabilistic loss
  function,'' in \emph{ICASSP}, Mar. 2020, pp. 56--60.

\bibitem{pratapMLSLargeScaleMultilingual2020}
V.~Pratap \emph{et~al.}, ``{MLS}: {A} large-scale multilingual dataset for
  speech research,'' in \emph{INTERSPEECH}, Oct. 2020, pp. 2757--2761.

\bibitem{heymannBeamnetEndtoendTraining2017}
J.~Heymann \emph{et~al.}, ``Beamnet: {End}-to-end training of a
  beamformer-supported multi-channel {ASR} system,'' in \emph{ICASSP}, Mar.
  2017, pp. 5325--5329.

\bibitem{Unified-Ochiai2017}
T.~Ochiai \emph{et~al.}, ``Unified architecture for multichannel end-to-end
  speech recognition with neural beamforming,'' \emph{IEEE J. Sel. Top. Signal
  Process.}, vol.~11, no.~8, pp. 1274--1288, 2017.

\bibitem{changMIMOSpeechEndtoendMultichannel2019}
X.~Chang \emph{et~al.}, ``{MIMO}-{Speech}: {End}-to-end multi-channel
  multi-speaker speech recognition,'' in \emph{ASRU}, Dec. 2019, pp. 237--244.

\bibitem{zhangEndtoEndFarFieldSpeech2020}
W.~Zhang \emph{et~al.}, ``End-to-end far-field speech recognition with unified
  dereverberation and beamforming,'' in \emph{INTERSPEECH}, Oct. 2020, pp.
  324--328.

\bibitem{zhangEndtoEndDereverberationBeamforming2021}
W.~Zhang~et al., ``End-to-end dereverberation, beamforming, and speech
  recognition with improved numerical stability and advanced frontend,'' in
  \emph{ICASSP}, Jun. 2021, pp. 6898--6902.

\bibitem{kimIndependentVectorAnalysis2006a}
T.~Kim \emph{et~al.}, ``Independent vector analysis: An extension of {ICA} to
  multivariate components,'' in \emph{International conference on independent
  component analysis and signal separation}.\hskip 1em plus 0.5em minus
  0.4em\relax Springer, 2006, pp. 165--172.

\bibitem{hiroeSolutionPermutationProblem2006}
A.~Hiroe, ``Solution of permutation problem in frequency domain {ICA}, using
  multivariate probability density functions,'' in \emph{{ASIACRYPT}
  2016}.\hskip 1em plus 0.5em minus 0.4em\relax Berlin, Heidelberg: Springer,
  Jan. 2006, vol. 3889, pp. 601--608.

\bibitem{onoStableFastUpdate2011}
N.~Ono, ``Stable and fast update rules for independent vector analysis based on
  auxiliary function technique,'' in \emph{WASPAA}, Oct. 2011, pp. 189--192.

\bibitem{scheiblerSurrogateSourceModel2021}
R.~Scheibler and M.~Togami, ``Surrogate source model learning for determined
  source separation,'' in \emph{ICASSP}, Jun. 2021, pp. 176--180.

\bibitem{boeddeker_comparison_2021}
C.~Boeddeker \emph{et~al.}, ``A comparison and combination of unsupervised
  blind source separation techniques,'' in \emph{Speech Communication; 14th ITG
  Conference}.\hskip 1em plus 0.5em minus 0.4em\relax VDE, 2021, pp. 1--5.

\bibitem{ilrma-t}
R.~Ikeshita \emph{et~al.}, ``A unifying framework for blind source separation
  based on a joint diagonalizability constraint,'' in \emph{EUSIPCO}, Sep.
  2019, pp. 1--5.

\bibitem{nakashimaJointDereverberationSeparation2021}
T.~Nakashima \emph{et~al.}, ``Joint dereverberation and separation with
  iterative source steering,'' in \emph{ICASSP}, Jun. 2021, pp. 216--220.

\bibitem{idlma}
N.~Makishima~et al., ``Independent deeply learned matrix analysis for
  determined audio source separation,'' \emph{IEEE/ACM Trans. Audio Speech
  Lang. Process.}, vol.~27, no.~10, pp. 1601--1615, 2019.

\bibitem{kameoka_supervised_2019}
H.~Kameoka \emph{et~al.}, ``Supervised determined source separation with
  multichannel variational autoencoder,'' \emph{Neural Computation}, vol.~31,
  no.~9, pp. 1891--1914, 09 2019.

\bibitem{saijo_tiss_2021}
K.~Saijo and R.~Scheibler, ``Low-memory end-to-end training for iterative joint
  speech dereverberation and separation with a neural source model,'' 2021.

\bibitem{duComputationallyEfficientOverdeterminedBlind2021}
Y.~Du \emph{et~al.}, ``Computationally-efficient overdetermined blind source
  separation based on iterative source steering,'' \emph{IEEE Signal Process.
  Lett.}, pp. 1--1, Dec. 2021.

\bibitem{Watanabe:2018gy}
S.~Watanabe~et al., ``{ESPnet}: End-to-end speech processing toolkit,'' in
  \emph{INTERSPEECH}, 2018, pp. 2207--2211.

\bibitem{karitaComparativeStudyTransformer2019}
S.~Karita~et al., ``A comparative study on transformer vs {RNN} in speech
  applications,'' in \emph{ASRU}, Dec. 2019, pp. 449--456.

\bibitem{Joint-Kim2017}
S.~Kim \emph{et~al.}, ``Joint {CTC}-attention based end-to-end speech
  recognition using multi-task learning,'' in \emph{ICASSP}, Mar. 2017, pp.
  4835--4839.

\bibitem{changEndToEndMultiSpeaker2020}
X.~Chang \emph{et~al.}, ``End-to-end multi-speaker speech recognition with
  transformer,'' in \emph{ICASSP}, 2020, pp. 6129--6133.

\bibitem{kolbaekMultitalkerSpeechSeparation2017}
M.~Kolbaek \emph{et~al.}, ``Multitalker speech separation with utterance-level
  permutation invariant training of deep recurrent neural networks,''
  \emph{IEEE/ACM Trans. Audio Speech Lang. Process.}, vol.~25, no.~10, pp.
  1901--1913, Aug. 2017.

\bibitem{nakataniSpeechDereverberationBased2010}
T.~Nakatani \emph{et~al.}, ``Speech {Dereverberation} {Based} on
  {Variance}-{Normalized} {Delayed} {Linear} {Prediction},'' \emph{IEEE Trans.
  Audio Speech Lang. Process.}, vol.~18, no.~7, pp. 1717--1731, Sep. 2010.

\bibitem{nakataniUnifiedConvolutionalBeamformer2019}
T.~Nakatani and K.~Kinoshita, ``A unified convolutional beamformer for
  simultaneous denoising and dereverberation,'' \emph{IEEE Signal Process.
  Lett.}, vol.~26, no.~6, pp. 903--907, Jun. 2019.

\bibitem{nakataniJointlyOptimalDenoising2020}
T.~Nakatani \emph{et~al.}, ``Jointly optimal denoising, dereverberation, and
  source separation,'' \emph{IEEE/ACM Trans. Audio Speech Lang. Process.},
  vol.~28, pp. 2267--2282, Jul. 2020.

\bibitem{scheibler_mm_2020}
R.~Scheibler and N.~Ono, ``{MM} algorithms for joint independent subspace
  analysis with application to blind single and multi-source extraction,''
  \emph{arXiv preprint arXiv:2004.03926}, 2020.

\bibitem{ikeshitaOverdeterminedIndependentVector2020}
R.~Ikeshita \emph{et~al.}, ``Overdetermined independent vector analysis,'' in
  \emph{ICASSP}, 2020, pp. 591--595.

\bibitem{togamiOverdeterminedSpeechSource2020a}
M.~Togami and R.~Scheibler, ``Over-determined speech source separation and
  dereverberation,'' in \emph{APSIPA}, Dec. 2020, pp. 705--710.

\bibitem{wsj1}
{Linguistic Data Consortium, and NIST Multimodal Information Group},
  \emph{{CSR-II} ({WSJ1}) Complete {LDC94S13A}}, Linguistic Data Consortium,
  Philadelphia, 1994, web Download.

\bibitem{barkerThirdCHiMESpeech2015}
J.~Barker \emph{et~al.}, ``The third ‘{CHiME}’ speech separation and
  recognition challenge: {Dataset}, task and baselines,'' in \emph{ASRU}, Nov.
  2015, pp. 504--511.

\bibitem{Mesaros2018_DCASE}
A.~Mesaros \emph{et~al.}, ``A multi-device dataset for urban acoustic scene
  classification,'' in \emph{DCASE}, November 2018, pp. 9--13.

\bibitem{paszkePyTorchImperativeStyle2019}
A.~Paszke~et al., ``{PyTorch}: An imperative style, high-performance deep
  learning library,'' in \emph{Advances in Neural Information Processing
  Systems}, H.~Wallach \emph{et~al.}, Eds., vol.~32.\hskip 1em plus 0.5em minus
  0.4em\relax Curran Associates, Inc., 2019.

\bibitem{scheiblerFastStableBlind2020}
R.~Scheibler and N.~Ono, ``Fast and stable blind source separation with rank-1
  updates,'' in \emph{ICASSP}, May 2020, pp. 236--240.

\bibitem{murataApproachBlindSource2001}
N.~Murata \emph{et~al.}, ``An approach to blind source separation based on
  temporal structure of speech signals,'' \emph{Neurocomputing}, vol.~41, no.
  1-4, pp. 1--24, Oct. 2001.

\bibitem{horiEndToEnd2018}
T.~Hori \emph{et~al.}, ``End-to-end speech recognition with word-based {RNN}
  language models,'' in \emph{Proc. IEEE SLT}, 2018, pp. 389--396.

\bibitem{koldovskyOrthogonallyConstrainedExtractionIndependent2018}
Z.~Koldovský \emph{et~al.}, ``Orthogonally-{Constrained} {Extraction} of
  {Independent} {Non}-{Gaussian} {Component} from {Non}-{Gaussian} {Background}
  {Without} {ICA},'' in \emph{Latent {Variable} {Analysis} and {Signal}
  {Separation}}.\hskip 1em plus 0.5em minus 0.4em\relax Cham: Springer, 2018,
  vol. 10891, pp. 161--170.

\end{thebibliography}

\end{document}